\documentclass[sigconf]{acmart}
\acmYear{2019}
\copyrightyear{2019} 
\setcopyright{rightsretained} 
\acmConference[ROME 2019]{Proceedings of the 1st ACM SIGIR Workshop on Reducing Online Misinformation Exposure}{July 25, 2019}{Paris, France}
\acmBooktitle{Proceedings of the 1st ACM SIGIR Workshop on Reducing Online Misinformation Exposure (ROME 2019), July 25, 2019, Paris, France}
\acmDOI{}
\acmISBN{}

\usepackage{algorithm}
\usepackage{algpseudocode}
\usepackage{float}
\usepackage{calrsfs}
\usepackage{lscape}
\usepackage{subfig}
\usepackage{graphicx}

\DeclareMathAlphabet{\pazocal}{OMS}{zplm}{m}{n}
\newcommand{\Ta}{\pazocal{T}}

\newcommand\norm[1]{\left\lVert#1\right\rVert}

\AtBeginDocument{%
  \providecommand\BibTeX{{%
    \normalfont B\kern-0.5em{\scshape i\kern-0.25em b}\kern-0.8em\TeX}}}

\begin{document}

%
\title{Tensor Factorization with Label Information \\for Fake News Detection}
%
%

\author{Frosso Papanastasiou}
\affiliation{%
  \institution{School of Electrical and Computer Engineering, National Technical University of Athens}
  \city{Athens}
  \country{Greece}}
\additionalaffiliation{%
  \institution{NCSR Demokritos}
  \city{Athens}
  \country{Greece}}

\email{efpanastasiou@gmail.com}

\author{Georgios Katsimpras}
\affiliation{%
  \institution{NCSR Demokritos}
  \city{Athens}
  \country{Greece}}
\email{gkatsibras@iit.demokritos.gr}

\author{Georgios Paliouras}
\affiliation{%
  \institution{NCSR Demokritos}
  \city{Athens}
  \country{Greece}}
\email{paliourg@iit.demokritos.gr}

%

%
\begin{abstract}
The buzz over the so-called ``fake news" has created concerns about a degenerated media environment and led to the need for technological solutions. As the detection of fake news is increasingly considered a technological problem, it has attracted considerable research. Most of these studies primarily focus on utilizing information extracted from textual news content. In contrast, we focus on detecting fake news solely based on structural information of social networks. We suggest that the underlying network connections of users that share fake news are discriminative enough to support the detection of fake news. Thereupon, we model each post as a network of friendship interactions and represent a collection of posts as a multidimensional tensor. Taking into account the available labeled data, we propose a tensor factorization method which associates the class labels of data samples with their latent representations. Specifically, we combine a classification error term with the standard factorization in a unified optimization process. Results on real-world datasets demonstrate that our proposed method is competitive against state-of-the-art methods by implementing an arguably simpler approach. 
\end{abstract}

%
%


%

%

%
\maketitle

\section{Introduction}
Many recent reports and studies suggest that the public is concerned about the impact of fake news. Specifically, a European-wide survey of 26,000 adults conducted during 2018, shows that fake news are widely spread across Europe and 85\% of the participants believe that fake news is a serious problem \cite{eusurvey}. The same study concludes that online social media is perceived as the least trusted source of news. Therefore, the need for a solution to stop the spread of fake news is emerging.

A great number of approaches to detecting fake news have been proposed in the literature. We can distinguish three main categories of such methods: i) content-based, ii) network-based and iii) hybrid. Content-based approaches emphasize on analysing the news content and trying to associate language patterns with deception \cite{hardalov2016search,rubin2016fake,Hosseinimotlagh2018}. Also called linguistic, these methods rely on Natural Language Processing (NLP) tasks such as syntactic, semantic and sentiment analysis. While these methods show good results, this is only achieved in closed domains where the context is already known and static \cite{granskogen2018automatic}. In contrast, network-based approaches focus on network information, such as friendship networks and propagation paths. These methods aim to extract information by analysing the connectivity of the networks \cite{gupta2012evaluating,jin2016news}. As such, network-based methods can perform well in dynamic environments and consequently in the task of fake news detection, which goes across topics and domains \cite{tacchini2017some}. Incorporating both content and network information, in a complementary way, is the goal of the hybrid approaches. Although this is a promising direction, combining together heterogeneous information is a hard and time-consuming task \cite{shu2017exploiting}. 

Based on the facts that i) users play an important role in the dissemination of fake news \cite{shu2017fake}, ii) users share information with similar users (friends) \cite{tian2012structural} and iii) network properties have been shown to be important for the classification of fake news \cite{shu2019studying}, we investigate relationship networks of users and how they facilitate the spreading of fake news.

Therefore, in this work, we focus on the information that can be extracted from structured networks and specifically from relationship networks between users. In particular, we model each post as a network of friendship interactions and represent a collection of posts as a multidimensional tensor. Using this representation we can employ tensor factorization methods to produce compact representations of posts that capture the underlying network information. However, the standard factorization methods have limited effectiveness due to their unsupervised nature. In cases, where labeled data are available, incorporating class information could assist the factorization process to produce discriminatory representations and thus to better identify fake news.  


\begin{sloppypar}
In this paper, we propose CLASS-CP, a tensor-based semi-supervised  approach for classifying fake news posts using network information and the available labeled data. Specifically, our approach combines tensor factorization and classification in a joint learning process. 
We model the data as a third-order tensor which represents the friendship relations that are present. At the same time, we use the labeled data we have in hand to influence the resulting factor matrices.    
To achieve this, we propose a tensor factorization method that assimilates class information about posts. While, commonly, factorization and classification are employed separately, we combine them in a single optimization process to obtain semi-supervised factorization. The main contributions of the paper are:
\end{sloppypar}
\begin{itemize}
    \item a tensor-based approach for detecting fake news that use network information and labeled data,
    \item combination of tensor factorization with classification, achieving class-driven modeling of tensorial data,
    \item evaluation of the performance of the method on real-world datasets, in order to demonstrate its effectiveness.
\end{itemize}

The code for the proposed method is publicly available and can be found at \url{https://github.com/FrossoPap/class-cp}.

\section{Related Work}



When it comes to fake news detection on social media, most approaches focus on finding patterns from news content, such as vocabulary, syntax, writing styles and images. For example, Hosseinimotlagh et al. \cite{Hosseinimotlagh2018} proposed an unsupervised tensor modeling of the problem, based on term frequency and spatial relations between terms and articles. Guacho et al. \cite{Guacho18} introduced a semi-supervised model via tensor embeddings that uses spatial/contextual information about news articles. Gupta et al. \cite{Gupta18} proposed a classifier to estimate tweet credibility from features such as number of words, URLs, hashtags, emojis, presence of swear words, pronouns. Horne et al. \cite{Horne17} employed an SVM-based algorithm using stylistic, complexity and psychological features of both title and body text to classify real, fake and satire news. However, most of these content-based methods require closed domains with predefined context in order to perform well. 

Recent work investigates network-based properties and features generated by the users' social profiles and interaction with the news. For example, Shu et al. \cite{Shu18} proposed to compare explicit and implicit user profile features, in order to measure their potential to identify fake news. Yang et al. \cite{Yang19} developed an unsupervised fake news detection algorithm to understand the dependencies among the truths of news, the users' opinions, and their credibility. Jin et al. \cite{Jin16} exploited conflicting viewpoints in news tweets with a topic model method and created  a credibility propagation network of tweets that generates the final result. Castillo et al. \cite{Castillo} exploited features from users' posting and re-posting behavior, text and citations to external sources. In another work, Shu et al. \cite{ShuWang18} proposed a tri-relationship embedding framework called TriFN, that models both publisher-news relations and user-news interactions for fake news classification. Existing state-of-the-art methods that consider network information, such as the TriFN framework, require the simultaneous modeling of many parameters that come from heterogeneous sources of information. This can be a hard and time consuming task.



%
In regard to the joint factorization framework that is proposed in this work, we can find several similar techniques in the literature. Context-aware methods \cite{Li10,Rendle11} combine contextual information with factorization to solve recommendation tasks. On the other hand, class-aware approaches \cite{cao2016semi,xiao2013class,katsimpras18} integrate class information into the factorization process but all these methods are optimized to perform in domains different from fake news detection. 

Our method builds on top of the class-aware approaches and aims to develop a less complex but still effective fake news detection system that exploits the network of friendship interactions and the class information of posts. The proposed method introduces an alternative way for modeling news posts through structural information and thus, it can be effective in numerous domains where content information is limited.


\section{Preliminaries and Notation}
Throughout the paper we use the following notation. The uppercase calligraphic letters denote tensors, e.g. $\Ta \in \mathbb{R}^{p\times u\times u}$, where $p$ is the number of posts and $u$ is the number of nodes (e.g. users). Matrices are represented by uppercase italic letters like \emph{A}. Lowercase bold letters, like $\mathbf{v}$, denote vectors. The \emph{(i,j)} element of a matrix  \emph{A} is denoted by $a_{ij}$. To refer to the \emph{i-th} row of a matrix  \emph{A} we use $\mathbf{a_i}$. Similarly, an element \emph{(i,j,k)} of a tensor $\Ta$ will be denoted as $\Ta_{ijk}$.
Additionally, $vec(\emph{A})$ is the vectorization of \emph{A}, the operator $\otimes$ is the Kronecker product and the operator $\odot$ is the Khatri-Rao product.
For our tensor modeling we use a binary representation: 
\[ \Ta_{ijk} =
  \begin{cases}
    1,       & \quad \text{if user k has engaged with post i}\\
    &\quad \text{and user j follows user k}\\
    0,  & \quad \text{otherwise}
  \end{cases}
\]

The \emph{order} of a tensor, also known as \emph{ways} or \emph{modes},  is the number of its dimensions, therefore, $\Ta$ is called a third-order tensor. Also, a $r$-order tensor is of rank-one if it can be strictly decomposed into the outer product of $r$ vectors. In addition, to simplify the notation, we follow the same notation as in \cite{kolda2006multilinear} for Kruskal operators, to express the factorization models in the next sections.


\section{Problem Definition and Proposed Method}
\label{sec_method}

\subsection{Problem definition}
Given a set of posts $P = (p_1,p_2,..p_p)$, a set of friendship networks $G = (g_1,g_2,..g_p)$ and assuming that a set of labels $Y = (y_1,y_2,..y_l)$ is available for some posts (e.g. $l<p$), we want to efficiently predict the labels of the remaining $p-l$ posts. Solving this task raises the following research questions:

\begin{itemize}
    \item (Q1) How can we combine the available friendship networks to facilitate the classification of nodes?
    \item (Q2) How can we incorporate the label information of the data in hand to enhance the analysis? 
\end{itemize}

\begin{figure*}[ht!]
\centerline{\includegraphics[width=0.8\textwidth]{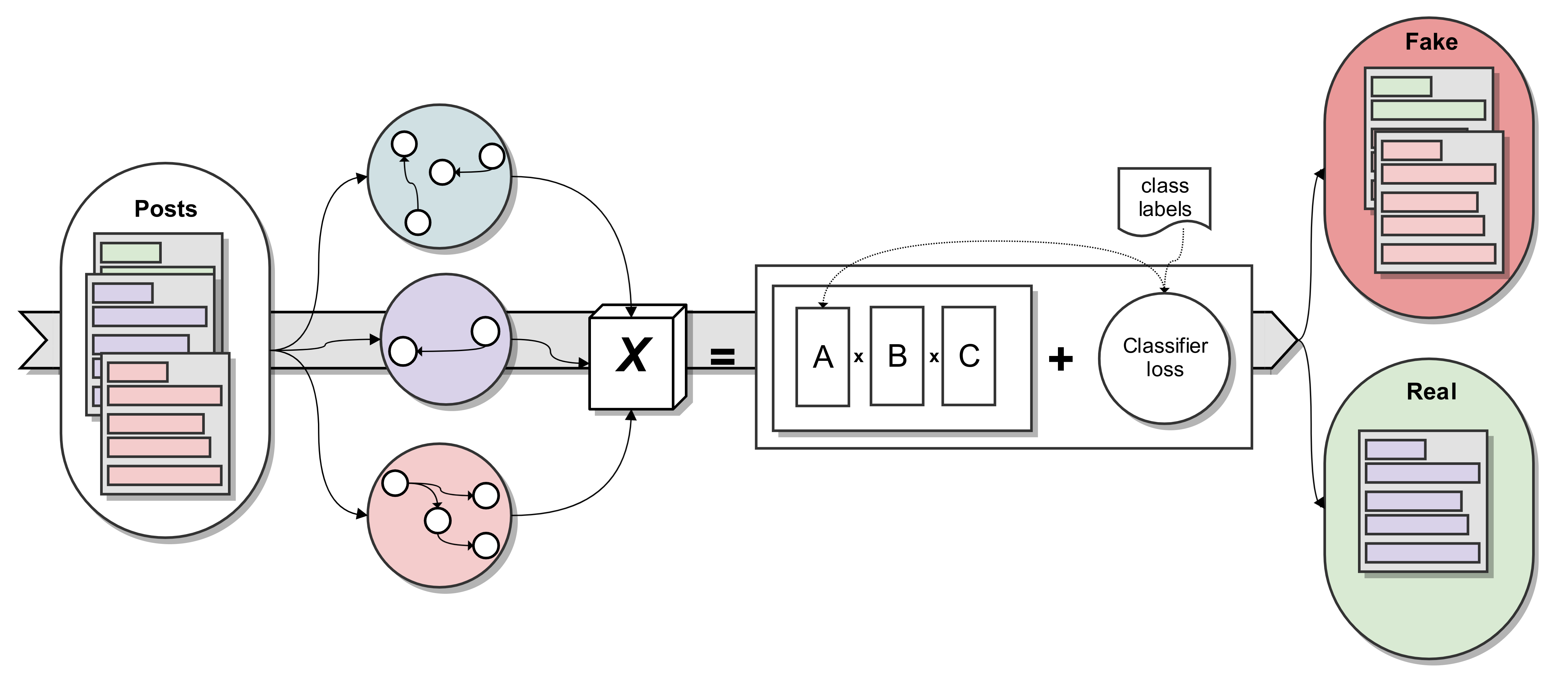}}
\caption{The conceptual diagram of CLASS-CP.}
\label{fig_flow}
\end{figure*}

\subsection{The CLASS-CP approach}

Our approach combines two basic mechanisms to solve questions (Q1) and (Q2): the collective learning achieved through tensor factorization and the class-aware optimization in relation to the latent factors, respectively. For the former task, we employ the Canonical/Parafac (CP) factorization \cite{Parafac}. CP is a widely used tensor factorization method, which decomposes a tensor into a sum of rank-one tensors. The first step in our approach is to build a tensor from the friendship networks, i.e. from social media data\footnote{It is worth noticing that the method can be readily applied to other domains as well (e.g. biology and medicine)}. The result is a 3rd-order tensor, $\Ta \in \mathbb{R}^{p\times u\times u}$. The CP decomposition of such a tensor is computed by the following least-squares loss:

\[ \text{min}_{A,B,C} \norm{\Ta - [\![A,B,C]\!]}^2  ,\] where $A \in \mathbb{R}^{p\times r}$ represents the posts-factor matrix, $B \in \mathbb{R}^{u\times r}$ and $C \in \mathbb{R}^{u\times r}$  represent the users-factor matrices, and $r$ is the rank of the factorization. 

To capture class-aware information we add a classification error term, based on the latent factor \emph{A}. Thus, CLASS-CP extends the CP factorization using supervision. That is, the 3rd-order tensor $\Ta$ is approximated, while taking into account the available labeled data. In particular, the post vectors in \emph{A}, are linked to class labels. 

The conceptual diagram of CLASS-CP is depicted in Figure \ref{fig_flow}. Firstly, we model posts as a network of friendship interactions and we represent a collection of them as a
three-dimensional tensor, though, CLASS-CP is able to accept higher modes as well. As we will show, in the learned latent space, the posts-factor matrix A is biased by a classifier. To achieve this, we employ a joint optimization process that learns to predict fake news posts.



\subsubsection{Joint optimization in CLASS-CP}
Assuming a set of labels for some of the posts, we expect similarly labeled posts to share similar factors\cite{mcpherson2001birds}. As mentioned above, we introduce class-label information into the tensor factorization, in order to move posts of the same class closer in the latent space. 
In particular, we introduce a matrix of coefficients $W \in \mathbb{R}^{r\times c}$, where $r$ is the number of factors and $c$ is the number of class labels. This matrix assigns labels to posts based on the factor matrix \emph{A} as in a common regression problem $Y=AW$.

Given a tensor $\Ta \in \mathbb{R}^{p\times u\times u}$ and a set of labels $Y \in \mathbb{R}^{l\times c}$, where $l$ is the number of labeled posts with $l<p$, and $c$ is the number of class labels, we solve the optimization problem presented in eq. \ref{eq_opt}.
\begin{equation}
\label{eq_opt}
\begin{aligned}
& \underset{A,B,C,W}{\text{minimize}}
&  f(A,B,C) + g(A,W), \\
\end{aligned}
\end{equation}
where 
 \[ f(A,B,C) = \norm{\Ta - [\![A,B,C]\!]}^2\] 
is the tensor factorization least squares problem,
\begin{equation}
\label{eq_knn}
\begin{aligned}
& g(A,W) = \lambda_g\norm{Y - DAW}^2
\end{aligned}
\end{equation}
 is the prediction error of the classifier
, where $D = [I^{lxl}, 0^{lx(p-l)}] \in \mathbb{R}^{l\times p}$ and  $\lambda_g$ is a hyperparameter to control the influence of the classification error in the optimization. Note that $g(A,W)$ is produced with respect to the current values of  the post factor matrix \emph{A}. 

We solve the minimization problem in eq. \ref{eq_opt}, using the efficient alternating least squares method (ALS). This approach alternately fixes and solves factor matrices following update rules. The update rules are derived by setting the gradient of eq. \ref{eq_opt} with respect to each factor matrix to zero.

For CP, the ALS method fixes every factor matrix except one and solves for it. The matricized form of \emph{f} (one per mode) can be written as:

\begin{equation}
\label{eq_mat_cp}
\begin{aligned}
& \Ta_{(1)} \approx A(C \odot B)^T ,\\
& \Ta_{(2)} \approx B(C \odot A)^T ,\\
& \Ta_{(3)} \approx C(B \odot A)^T .
\end{aligned}
\end{equation}
Recall that the $\odot$ operator denotes the Khatri-Rao product.

\subsubsection{Updating factor matrices B and C}
To find the updates for \emph{B} and \emph{C}, we ignore the second term of eq.\ref{eq_opt} and we solve the respective equations as presented in eq. \ref{eq_mat_cp}. The ALS approach fixes \emph{A} and \emph{C} to solve for B, and \emph{A}, \emph{B} to solve for C. Let $Z_{(2)}=(C \odot A)$ and $Z_{(3)}=(B \odot A)$, then by combining eq. \ref{eq_opt} and eq. \ref{eq_mat_cp} we can write:
\begin{equation}
\label{eq_b}
\begin{aligned}
& \underset{B}{\text{min}}
&  \norm{\Ta_{(2)} - BZ_{(2)}^T}^2 \\
\end{aligned}
\end{equation}
and
\begin{equation}
\label{eq_c}
\begin{aligned}
& \underset{C}{\text{min}}
&  \norm{\Ta_{(3)} - CZ_{(3)}^T}^2. \\
\end{aligned}
\end{equation}
The solution to eq. \ref{eq_b} and eq. \ref{eq_c} is given by \cite{kolda2009tensor} and so the updates of \emph{B} and \emph{C} can be written as:
\begin{equation}
\label{sol_b}
\begin{aligned}
& B = (Z_{(2)}^TZ_{(2)})^{-1}Z_{(2)}^T\Ta_{(2)} \\
\end{aligned}
\end{equation}

and

\begin{equation}
\label{sol_c}
\begin{aligned}
& C = (Z_{(3)}^TZ_{(3)})^{-1}Z_{(3)}^T\Ta_{(3)} \\
\end{aligned}
\end{equation}

\subsubsection{Updating factor matrix A}
Let $Z_{(1)}=(C \odot B)$. To find the update for factor matrix \emph{A}, we can combine eq. \ref{eq_opt} and eq. \ref{eq_mat_cp} and write:

\begin{equation}
\label{eq_a}
\begin{aligned}
& \underset{A}{\text{min}}
&  \norm{\Ta_{(1)} - AZ_{(1)}^T}^2 + \norm{Y - DAW}^2 .\\
\end{aligned}
\end{equation}

Equation \ref{eq_a} can be solved through its vectorized form:

\begin{equation}
\label{eq_a_vec}
\begin{split}
& \underset{A}{\text{min}}\quad
\norm{vec(\Ta_{(1)}) - (Z_{(1)}\otimes I_p)vec(A)}^2 + \\
&  \quad \quad \norm{vec(Y) - (W^T\otimes D)vec(A)}^2 .
\end{split}
\end{equation}
Letting $G=(Z_{(1)} \otimes I_p)$ and  $L=(W^T \otimes D)$, the solution to eq. \ref{eq_a_vec} can be calculated and the update for \emph{A} is given by:

\begin{equation}
\label{sol_a}
\begin{aligned}
& vec(A) = (G^TG + L^TL)^{-1}(G^Tvec(\Ta_{(1)}) + L^Tvec(Y)) \\
\end{aligned}
\end{equation}

\subsubsection{Updating factor matrix W}
To find the update of \emph{W}, we ignore the first term of eq. \ref{eq_opt} and so we solve:
\begin{equation}
\label{eq_w}
\begin{aligned}
& \underset{W}{\text{min}}
&  \norm{Y - DAW}^2 .\\
\end{aligned}
\end{equation}
The solution of eq. \ref{eq_w} with respect to \emph{W} gives the update of \emph{W}:
\begin{equation}
\label{sol_w}
\begin{aligned}
& W = (A^TD^TDA)^{-1}A^TD^TY \\
\end{aligned}
\end{equation}

\begin{algorithm}
\caption{CLASS-CP: Given a tensor $\Ta$ and a set of labels Y, approximate A,B,C and W}
\label{alg_re}
\begin{algorithmic}[1]
\Require{tensor $\Ta$, labels Y }
\Ensure{factor matrices A,B,C and coefficients W}
\State Initialize A,B,C,W and hyperparameter $\lambda_g$
\Repeat \\
\quad update{A} using Eq. \ref{sol_a} \\
\quad update{B} using Eq. \ref{sol_b} \\
\quad update{C} using Eq. \ref{sol_c} \\
\quad update{W} using Eq. \ref{sol_w}
\Until convergence

\end{algorithmic}
\end{algorithm}

The algorithm of CLASS-CP that brings together all of the above is depicted in Alg. \ref{alg_re}. To compute the factor matrices \emph{A}, \emph{B}, \emph{C} and the coefficient matrix \emph{W} the algorithm performs alternating updates until it converges to a criterion or it reaches a maximum number of iterations. As a convergence criterion we use the \emph{relative change} which can be calculated in each iteration as:
\begin{equation}
\label{criterion}
\begin{aligned}
& \frac{|(f_{new} + g_{new}) - (f_{old} + g_{old})|}{f_{old} + g_{old}} \\
\end{aligned}
\end{equation}


\section{Experimental Evaluation}
\subsection{Datasets}
\begin{table}
  \caption{Numbers of FakeNewsNet dataset}
  \label{tab:data}
  \begin{tabular}{ccl}
    \toprule
    Media Platform & BuzzFeed & PolitiFact\\
    \midrule
    \# Users & 1449 & 1697\\
    \# Engagements & 8598 & 10249\\
    \# Social Links & 6571 & 3093\\
    \# True news & 91 & 120\\
    \# Fake news & 91 & 120\\
  \bottomrule
\end{tabular}
\end{table}
To evaluate our approach we used two real-world public datasets which have been used previously in the literature \cite{shu2017exploiting, shu2017fake, shu2018fakenewsnet}. The datasets were collected from two platforms, BuzzFeed and PolitiFact. They include both news content and social context features with fact-checked ground truth labels. News content features include meta-information such as the title and body text, and social context includes the related user's profile information and activity (e.g., user sharing news on Twitter, user's Twitter followers). For our evaluation, we use only social context features such as the friendship network which indicates the following/followee structure of users who post related posts.

In order to decrease the size and the sparsity of the data, we removed users with a node degree < 3. As a result, we produced two tensors of size 182x1449x1449 and 240x1697x1697 for the BuzzFeed and PolitiFact datasets respectively. The basic statistics of the datasets after nodes deletion are shown in more detail in Table \ref{tab:data}. User engagements include posts, re-posts and replies related with a news article.

\subsection{Experimental Settings}

To calculate the performance of CLASS-CP, we use  Accuracy, Precision, Recall, and F1, that are commonly used as evaluation metrics in similar problems. We choose the first 80\% of news posts as our training set and the remaining 20\% for testing. Our data is balanced, meaning that the number of news posts labeled as true is equal to the number of fake posts. We perform the same process 10 times, independently for each dataset, and we report the average results.

\begin{figure*}[ht!]
    \centering
    \subfloat[PolitiFact]{\includegraphics[width=0.45\linewidth]{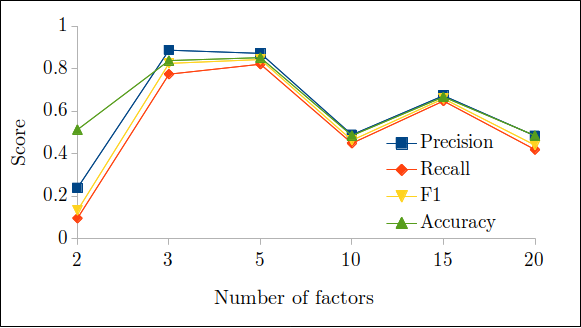}}%
    \qquad
    \subfloat[BuzzFeed]{\includegraphics[width=0.45\linewidth]{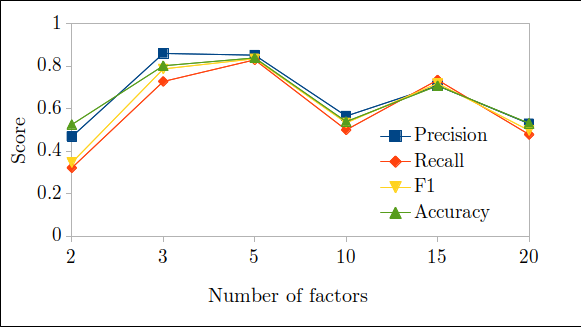}}%
    \caption{Behavior of CLASS-CP with different number of factors for each dataset.}%
    \label{fig:factors}%
\end{figure*}

\begin{figure}[ht!]
    \centering
    \subfloat{\includegraphics[width=0.5\textwidth]{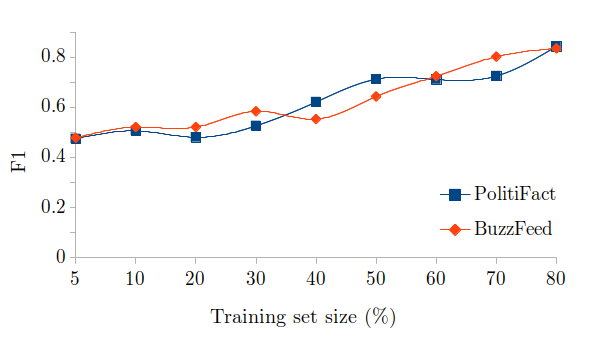}}%
    \qquad
    \caption{The learning curve on BuzzFeed and PolitiFact}%
    \label{fig:lc}%
\end{figure}

\textbf{Number of factors.} Before comparing CLASS-CP against other approaches, we measure the effect of the number of factors on the results. We set the number of factors r = 2, 3, 5, 10, 15, 20. A larger r increases the computation time of the factorization. The best overall performance is achieved for r = 5 as shown in Figure \ref{fig:factors} and we use it for the rest of the experiments.

\textbf{Impact of training set size on performance.} Additionally, we explore the behavior of CLASS-CP with respect to the size of the labeled data. In Figure \ref{fig:lc}, we show the F1-score learning curves in relation to the labeled data ratio on both datasets. From the plot, we can see that the more labeled instances are used in the training stage, the better performance.

\textbf{Hyperparameter $\lambda_g$.} We set the value of the hyperparameter $\lambda_g=1$ so as the classification loss can contribute equally to the optimization process, as the factorization.

\textbf{Execution time.} Each iteration takes less than a minute to finish and we need on average a few minutes ($<$10) to reach convergence.

\subsection{Classification of Fake News}

We compare the proposed CLASS-CP algorithm with the following methods:

\begin{itemize}
    \item \textbf{SVM:} 
    A baseline linear SVM model that is built by using the rows of the tensor $\Ta$ as input feature vectors.
    \item \textbf{CP+SVM:} The original CP factorization combined with a linear SVM classifier. This is a two-step process. In the first step we calculate the tensor embeddings of posts using CP in an unsupervised setting and in the second step we train a SVM classifier using the embeddings produced in the first step.
    \item \textbf{CP+\emph{k}-NN:} The CP factorization combined with a \emph{k}-NN classifier. We follow the same procedure as in CP+SVM but instead of a SVM classifier, we use a \emph{k}-NN classifier. As suggested by \cite{Duda, Hassanat}, we set \emph{k} equal to the square root of the size of the training data i.e. $k = \sqrt{p}$.   
    \item \textbf{TriFN:} TriFN is a state-of-the-art fake news detection method that extracts features separately from news publisher and user interactions and captures the interrelationship simultaneously. We use the same evaluation settings as proposed in \cite{ShuWang18} 
    so as to compare our results with the ones reported in that study. 
    
\end{itemize}

\begin{figure}[ht!]
    \centering
    \subfloat[PolitiFact]{\includegraphics[width=0.45\textwidth]{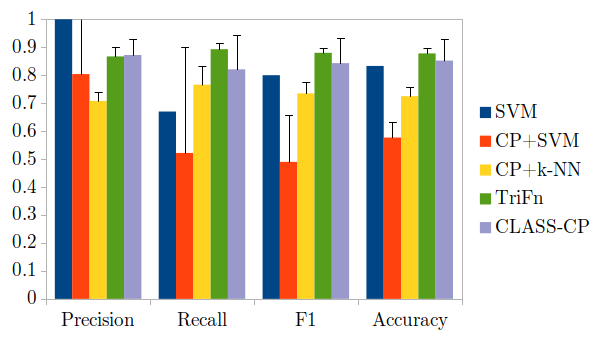}}%
    \qquad
    \subfloat[BuzzFeed]{\includegraphics[width=0.45\textwidth]{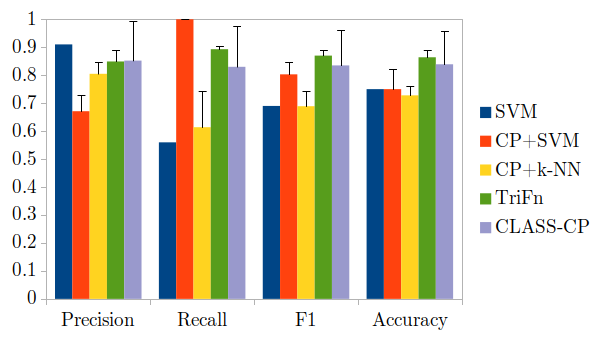}}%
    \caption{Performance comparison for different metrics}%
    \label{fig:plot}%
\end{figure}

\textbf{Experimental results.} Figure \ref{fig:plot} and Tables \ref{tab:comppol} and \ref{tab:compbuzz} show the comparison of CLASS-CP and the aforementioned methods for different metrics. From these, we can make the following observations: 

\begin{itemize}
    \item The SVM baseline approach gives the highest Precision results, but is weaker in terms of the other metrics. 
    \item The use of CP does not improve the performance of the SVM.
    \item CLASS-CP significantly outperforms SVM, CP+SVM and CP+ \emph{k}-NN both on the PolitiFact and the BuzzFeed datasets. 
    \item CLASS-CP performs comparable to TriFN in both datasets. 
\end{itemize}

\begin{table*}[ht]
\caption{Performance comparison on PolitiFact}
\label{tab:comppol}
\begin{tabular}{|l|l|l|l|l|l|}
\hline
Metric & SVM & CP+SVM & CP+ \emph{k}-NN & TriFN & CLASS-CP \\ \hline
Precision & 1.000 & $0.804 \pm 0.253$ & $0.708 \pm 0.033$ & $0.867 \pm 0.034$ & $0.872 \pm 0.058$ \\ \hline
Recall & 0.670 & $0.522 \pm 0.380$ & $0.766 \pm 0.066$ & $0.893 \pm 0.023$ & $0.821 \pm 0.122$  \\ \hline
F1 & 0.800 & $0.490 \pm 0.168$ & $0.735 \pm 0.039$ & $0.880 \pm 0.017$ & $0.843 \pm  0.089$  \\ \hline
Accuracy & 0.833 & $0.577 \pm 0.055$ & $0.725 \pm 0.034$ & $0.878 \pm 0.020 $& $0.852 \pm 0.078$ \\ \hline
\end{tabular}
\end{table*}

\begin{table*}[ht]
\caption{Performance comparison on BuzzFeed}
\label{tab:compbuzz}
\begin{tabular}{|l|l|l|l|l|l|}
\hline
Metric & SVM & CP+SVM & CP+ \emph{k}-NN & TriFN & CLASS-CP \\ \hline
Precision & 0.910 & $0.671 \pm 0.058$ & $0.805 \pm 0.041$ &$0.849 \pm 0.040$ & $0.852 \pm 0.143$ \\ \hline
Recall & 0.560 & 1.000 & $0.614 \pm 0.137$ & $0.893 \pm 0.013$ & $0.830 \pm 0.146$ \\ \hline
F1 & 0.690 & $0.803 \pm 0.044$ & $0.689 \pm 0.056$ & $0.870 \pm 0.019$ & $0.835 \pm 0.127$ \\ \hline
Accuracy & 0.750 & $0.750 \pm 0.071$ & $0.728 \pm 0.032$ & $0.864 \pm 0.026$ & $0.839 \pm 0.118$ \\ \hline
\end{tabular}
\end{table*}

The above suggest that (i) using only network information and the available labeled data, (ii) by incorporating class information during and not after the factorization process and (iii) with a small number of factors that lead to short computation times, we can achieve an equally good performance as state-of-the-art frameworks, such as the TriFN, that require the fusion of many heterogeneous data sources and complex calculations. Therefore, we confirm our initial intuition that the use of the underlying network between users can reveal valuable information about fake news.

\section{Conclusion}

In this paper, instead of employing tensor factorization and classification separately, as is common, we propose a method that combines them in a joint learning process for detecting fake news posts in social media. This tensor-based approach uses the follower/followee structure of users that have engaged with the news, as well as a set of labeled news that are provided. In order to combine this information, we introduced an extension of the standard tensor factorization method CP that incorporates class-label information into the factorization itself. In this manner we obtain a class-aware semi-supervised tensor factorization method.

For the evaluation of the method, we conducted experiments with two real-world public datasets that confirm the importance of incorporating the class information in the factorization process, rather than using it in a separate step. The results also confirmed our initial intuition, that the way users are connected with each other can reveal the truthfulness of the news they share on social media.

As further work, we would like to explore the extent to which CLASS-CP improves when more features, both from the network and the content of the posts, are added. Additionally, we plan to assess the performance of our approach on more datasets and investigate the impact of data size on scalability. At a methodological level, it would be interesting to explore new ways of representing the available information with tensors, in order to incorporate it in the proposed framework.

\bibliographystyle{ACM-Reference-Format}
\bibliography{sample-base}

%
\appendix

\end{document}